\pdfoutput=1
\def\preprint{0} % Define to 0 for two-column layout

\if\preprint0 % Submission setup
\documentclass[prl,twocolumn,superscriptaddress,showkeys,floatfix,10pt]{revtex4-2}
\else % Preprint setup
\documentclass[preprint,superscriptaddress,showkeys,floatfix,12pt]{revtex4-2}
\usepackage{lineno}\linenumbers
\fi

\usepackage[utf8]{inputenc}           % UTF-8 characters
\usepackage[english]{babel}           % English
\usepackage{hyperref}                 % Hyperlinks
\usepackage{siunitx}                  % Printing numbers and units with \SI{}{}
\usepackage{graphicx}                 % Figures
\usepackage{microtype}                % Better kerning
\usepackage{natbib}                   % Better citing and references
\usepackage{xcolor}                   % Colors
\usepackage{tikz}                     % Drawing geometric shapes
\usepackage{booktabs}                 % Better tables
\usepackage{newtxtext}                % Font that looks closer to Times new roman
\usepackage{newtxmath}                % Font for maths
\usepackage[capitalize]{cleveref}     % Smart figure references with \cref{}/\Cref{}
\usepackage{placeins}                 % For \FloatBarrier
\usepackage[normalem]{ulem}           % Strike-through text

\graphicspath{{figs/}}                % Implicitly look for figures in figs
\usetikzlibrary{calc}                 % Coordinate calculations
\definecolor{links}{RGB}{56,115,178}  % Blue color for link borders
\hypersetup{allbordercolors=links}    % Color all link borders in blue

\begin{document}

\newcommand{\hlnote}[1]{\textcolor{red!60!black}{#1}}
\newcommand{\added}[1]{\textcolor{green!60!black}{#1}}
\newcommand{\removed}[1]{\textcolor{red!60!black}{\sout{#1}}}

\title{From molecular to multi-asperity contacts: how roughness bridges the friction scale gap}

\author{Lucas Frérot}
\affiliation{Department of Physics and Astronomy, Johns Hopkins University, 3400 N Charles Street, Baltimore, Maryland 21218, USA}
\affiliation{Department of Mechanical Engineering, Johns Hopkins University, 3400 N Charles Street, Baltimore, Maryland 21218, USA}

\author{Alexia Crespo}
\affiliation{Laboratoire de Tribologie et Dynamique des Systèmes, École Centrale de Lyon, CNRS UMR5513, 69134, Ecully, France}

\author{Jaafar A. El-Awady}
\affiliation{Department of Mechanical Engineering, Johns Hopkins University, 3400 N Charles Street, Baltimore, Maryland 21218, USA}

\author{Mark O. Robbins}
%\thanks{Deceased 2020-08-13}
\affiliation{Department of Physics and Astronomy, Johns Hopkins University, 3400 N Charles Street, Baltimore, Maryland 21218, USA}

\author{Juliette Cayer-Barrioz}
\affiliation{Laboratoire de Tribologie et Dynamique des Systèmes, École Centrale de Lyon, CNRS UMR5513, 69134, Ecully, France}

\author{Denis Mazuyer}
\affiliation{Laboratoire de Tribologie et Dynamique des Systèmes, École Centrale de Lyon, CNRS UMR5513, 69134, Ecully, France}

\date{\today}

% PDF Metadata
\newcommand{\mykeywords}{friction; transient response; roughness; contact junction; fatty acid monolayers}
\makeatletter
\hypersetup{%
    pdftitle={\@title},
    pdfauthor={Lucas Frérot, Alexia Crespo, Jaafar A. El-Awady, Mark O. Robbins, Juliette Cayer-Barrioz, Denis Mazuyer},
    pdfkeywords={\mykeywords}
}
\makeatother

%%%%%%%%%%%%%%%%%%%%%%%%%%%%%%%%%%%%%%%%%%%%%%%%%%%%%%%%%%%%%%%%%%%%%%%%%%%%%%%%

\begin{abstract}
Friction is a pervasive phenomenon that affects the mechanical response of natural and man-made systems alike, from vascular catheterization to geological faults responsible for earthquakes. While friction stems from the fundamental interactions between atoms at a contact interface, its best descriptions at the macroscopic scale remain phenomenological. The so called ``rate-and-state'' models, which specify the friction response in terms of the relative sliding velocity and the ``age'' of the contact interface, fail to uncover the nano-scale mechanisms governing the macro-scale response, while models of friction at the atomic scale often overlook how roughness can alter the friction behavior. Here we bridge this gap between nano and macro descriptions of friction by correlating the physical origin of macroscopic friction to the existence, due to nanometric roughness, of contact junctions between adsorbed monolayers. Their dynamics, as we show, emerges from molecular motion. Through coupled experimental and atomic simulations, we highlight that transient friction overshoots its steady-state value after the system is allowed to rest, with the friction force decaying to a steady-state value over a distance of a few nanometers, much smaller than the junction size, even with a root-mean-square roughness of \SI{0.6}{\nm}. We demonstrate how this transient decay is intrinsically related to the evolution of the number of cross-surface attractive physical links between adsorbed molecules on rough surfaces. We also show that roughness is a sufficient condition for the appearance of frictional aging. In systems that show structural aging, this paints contact junctions as a key component in the observation of the transient friction overshoot, and shows how infrajunction molecular motion can control the macroscopic response. This is further corroborated by a multi-scale---in both time and space---theoretical approach, which accurately reproduces the transient friction response. Our results demonstrate that understanding how the friction mechanisms within contact junctions interact with nanometric roughness is instrumental in lifting the phenomenological veil over macro-scale friction models. Following the multi-scale framework we propose, enriching friction models to include scale interactions should have a broad impact on the description of friction in many systems, from the slip of glaciers to synovial joints.

\end{abstract}


%%%%%%%%%%%%%%%%%%%%%%%%%%%%%%%%%%%%%%%%%%%%%%%%%%%%%%%%%%%%%%%%%%%%%%%%%%%%%%%%

\maketitle

Friction is a phenomenon that affects the behavior of virtually every mechanical system: from the movement of geological faults that can cause earthquakes to the sliding of atomic-force microscopy tips. In these systems, friction arises from the interaction of contacting asperities~\citep{bowdenAreaContactStationary1939,bowdenMechanismMetallicFriction1942}: the inevitable roughness of natural and manufactured surfaces implies that the true contact interface is made up of a sparse set of contacts junctions~\citep{dieterichDirectObservationFrictional1994} which governs the frictional response~\citep{bowdenMechanismMetallicFriction1942,ben-davidStaticFrictionCoefficient2011}, as well as other tribological phenomena~\citep{archardContactRubbingFlat1953,greenwoodContactNominallyFlat1966,frerotMechanisticUnderstandingWear2018}.

At macroscopic scales, the static friction force has been observed to increase logarithmically with resting contact time for amorphous materials, including woods~\citep{coulombTheorieMachinesSimples1821}, rocks~\citep{dieterichModelingRockFriction1979} and polymers~\citep{dieterichDirectObservationFrictional1994}. This is attributed to an increase of the true contact area due to a mechanical creeping of contact spots~\citep{dieterichDirectObservationFrictional1994} (geometrical aging), an increase of the number of chemical bonds between the two surfaces~\citep{liFrictionalAgeingInterfacial2011,tianMemoryDistanceInterfacial2019} (chemical aging), or increase in interaction energy between contacting surfaces~\citep{bureauElasticiteRheologieInterface2002,corwinFrictionalAgingSliding2009} (structural aging). Upon sliding, the contact interface rejuvenates over a characteristic sliding distance $D_0$~\citep{ruinaSlipInstabilityState1983,georgesInterfacialFrictionWetted1994}. 
Such behavior is widely modeled using rate-and-state friction~\citep{dieterichModelingRockFriction1979,riceStabilitySteadyFrictional1983,brenerUnstableSlipPulses2018}, which describes the friction force in terms of a state variable $\phi$, that represents the average age of the micro-contacts and whose evolution equation encompasses aging and rejuvenation.
Despite recent efforts to relate rate-and-state parameters to the physics of rough surfaces~\citep{estrinModelFrictionalSliding1996,baumbergerSolidFrictionStick2006,aharonovPhysicsBasedRockFriction2018,molinariFundamentalAspectsNew2019,tianMemoryDistanceInterfacial2019}, in practice $D_0$ remains a phenomenological variable fitted to laboratory experiments. 

At nanoscopic scales, friction stems from inter-atomic forces between the surfaces in contact. While the effects of adsorbed layers~\citep{heAdsorbedLayersOrigin1999,mazuyerFrictionDynamicsConfined2008}, disorder~\citep{montiSlidingFrictionAmorphous2020}, lattice commensurability~\citep{hodStructuralSuperlubricityUltralow2018} are well known at the nano-scale, roughness at these scales can still break the contact area down into small junctions whose collective behavior may be different from a perfectly smooth response.

Our aim here is therefore two-fold: elucidating the influence of roughness on nano-scale friction mechanisms and integrating the physical contribution of these mechanisms into a macroscopic friction description. To do this, we focus on a representative model system of two rough cobalt surfaces coated with a stearic acid (C\textsubscript{17}H\textsubscript{35}COOH, commonly used as an environmentally-friendly lubricant) in dodecane (C\textsubscript{12}H\textsubscript{26}) dilute solution. After deposition of the solution, the stearic acid adsorbs on the surfaces and forms a monolayer~\citep{crespoEffectUnsaturationAdsorption2018,abouhadidFrictionLawsSaturated2021}. These two rough monolayer-covered surfaces are brought into contact in our molecular tribometer~\citep{crespoMethodologyCharacterizeRheology2017} at constant normal force. A slide-hold-slide protocol is applied with constant velocity and varied hold times. Molecular dynamics (MD) simulations reproducing the experimental protocol (at shorter timescales) are used to probe the details of the contact interface, for which we combine the two surfaces roughness profiles into a single rough-on-flat setting (roughness profiles are generated using measurements of the experimentally-used surfaces). Nano-scale mechanisms uncovered with MD and experimental results are used to establish a unifying friction model that we show reproduces the transient friction behavior observed in experiments.

\section*{Frictional aging and transient overshoot}

\begin{figure*}
% Vertically centering the two figures with \raisebox
\raisebox{-.5\height}{\includegraphics{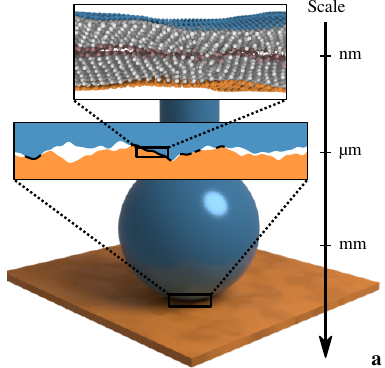}}
\raisebox{-.5\height}{\includegraphics{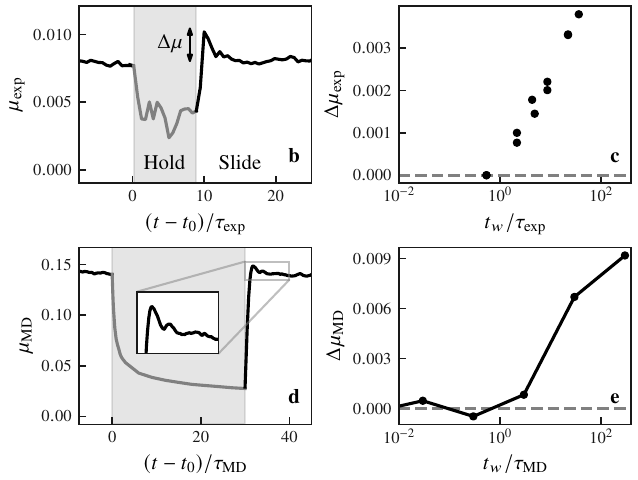}}
\caption{\textbf{Transient friction behavior of stearic acid monolayers.} Schematic (\textbf{a}) of a ball-on-flat contact experiment for fatty-acid monolayers showing the multi-scale nature of friction. The apparent contact area has a radius of \SI{2.45}{\um}, while the true contact area is made up of sparse junctions where the adsorbed monolayers interact due to the surface roughness. The friction response of the experiments ($\pm$ 10\% error) and simulations are shown in \textbf{b}--\textbf{c} and \textbf{d}--\textbf{e}, respectively. \textbf{b} and \textbf{d} show the transient friction behavior in a slide-hold-slide protocol, with hold highlighted in gray ($t_0$ being the start time of the hold stage). An overshoot of the steady-state friction force can be observed at the onset of the second slide stage. \textbf{c} and \textbf{e} show that the magnitude of the overshoot increases with the hold time, $t_w$, if it is greater than a relaxation time of $\tau_\mathrm{exp}$ = \SI{2.2}{\s} in the experiment and $\tau_\mathrm{MD}$ = \SI{0.8}{\ns} in the simulations.}\label{fig:observation}
\end{figure*}

\Cref{fig:observation}a illustrates the multi-scale aspect of friction of surfaces coated with fatty acid monolayers: the inevitable roughness of the surfaces in contact partitions the apparent contact interface into contact junctions~\citep{dieterichDirectObservationFrictional1994} where the fatty acid molecules are close enough to interact. This occurs, as we show in this work, even with a root-mean-square (RMS) roughness as small as \SI{0.6}{\nm}, as measured in the current experiments with atomic force microscopy (AFM) over \SI{1}{\um^2}. Figures \ref{fig:observation}b and \ref{fig:observation}d show the transient friction response of stearic acid monolayers for a slide-hold-slide (SHS) protocol, where $t_0$ is the start time of the holding stage, $\mu_\mathrm{exp}$ and $\mu_\mathrm{MD}$ are the ratios of tangential to normal force for the experiment and simulation, respectively. During the holding stage, relaxation causes the friction force to decrease to a non-zero value. After rest, the friction force overshoots the steady-state value by $\Delta\mu_\mathrm{exp}$ (resp.\ $\Delta\mu_\mathrm{MD}$). This overshoot is observed in both the experiments and MD simulations of rough-on-flat (\emph{cf}.\ \cref{fig:observation}d) and rough-on-rough (\emph{cf}.\ supplementary \cref{ex:rough_static_peaks}), and is consistent with previous observations of frictional aging in experiments~\citep{coulombTheorieMachinesSimples1821,dieterichModelingRockFriction1979,ruinaSlipInstabilityState1983,corwinFrictionalAgingSliding2009} and simulations~\citep{liFrictionalAgeingInterfacial2011} at a macroscopic scale. Figures \ref{fig:observation}c and \ref{fig:observation}d show that for both the experiments and the simulations the magnitude of the overshoot increases with the waiting time, $t_w$, for times longer than the relaxation times $\tau_\mathrm{exp}$ = \SI{2.2}{\s} and $\tau_\mathrm{MD}$ = \SI{0.8}{\ns} of the experiment and simulation, respectively. We have defined $\tau_\mathrm{exp}$ directly from \cref{fig:observation}c, but $\tau_\mathrm{MD}$ is defined from the mean-square displacement of monomers in an equilibrium simulation (see supplementary \cref{ex:msqd}), hence, our interpretation of $\tau_\mathrm{exp}$ and $\tau_\mathrm{MD}$ as relaxation time-scales. Independent simultaneous measurement of the tangential stiffness in the SHS experiment~\citep{crespoComprehensionTribologieFilms2017} shows a reversible increase of the stiffness during the hold step (see supplementary \cref{ex:tangential_stiffness}). This, combined with previous observation of the decreasing film thickness at rest~\citep{georgesInterfacialFrictionWetted1994}, confirms the presence of structural aging at the molecular scale during rest. With both our experimental and computational systems showing evidence of aging, we investigate the role of roughness in the observed response.

\begin{figure*}
\includegraphics{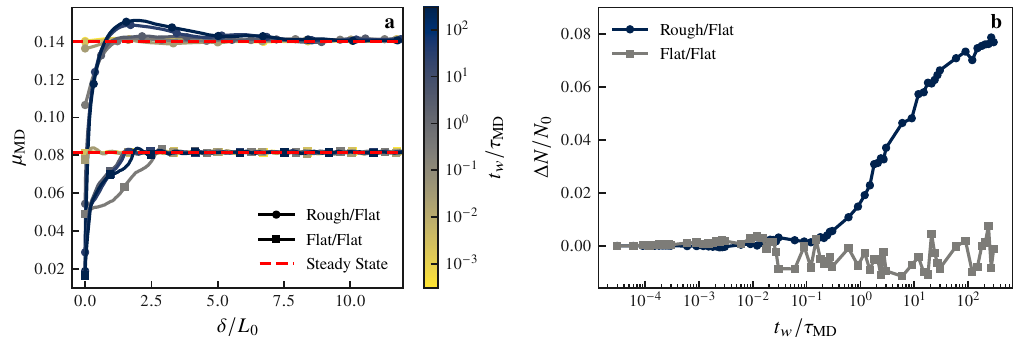}
\caption{\textbf{Effects of roughness on the transient friction response}. \textbf{a} compares the transient friction response of a rough-on-flat and a flat-on-flat system after a hold time $t_w$ (darker curves have longer $t_w$). The flat/flat system shows that the friction force recovers a steady-state value without overshooting, unlike the rough/flat system, which exhibits a friction force peak above $\mu_\mathrm{ss}$ for large $t_w$. \textbf{b} compares the increase in the number of cross-surface links ($\Delta N$) in the holding stage. While $\Delta N$ increases markedly for the rough/flat system, it remains constant in the flat/flat system.}\label{fig:roughness}
\end{figure*}

\section*{Role of roughness in aging}
We compare in \cref{fig:roughness}a the MD transient friction response (as a function of sliding distance, $\delta$, normalized by the molecule length, $L_0$ = \SI{2.14}{\nm}) of a flat-on-flat system with a rough-on-flat system, as well as the increase in number of cross-surface links (XSLs), during the hold step, i.e.\ the number of attractive interactions, due to van der Waals forces, between the atoms of the molecules adsorbed on opposite surfaces. In \cref{fig:roughness}a, the rough-on-flat system exhibits a friction overshoot as showcased in \cref{fig:observation}, while the flat-on-flat system does not. Instead, the transient of the latter consists in an increase of the friction force only up to the steady-state value, regardless of $t_w$. Roughness, even at this nanometric scale, is therefore a sufficient condition for the occurrence of an aging process. In order to understand the fundamental mechanisms underlying the transient friction response, we look at the number of XSLs, $N$, and how it increases relative to the value at the end of the initial sliding phase, $N_0$. The change of number of links, $\Delta N = N - N_0$, indicates structural aging because the contact area is constant, and no covalent bonds are formed (see supplementary \cref{ex:contact_map_wait}). \Cref{fig:roughness}b shows that the number of XSLs in the flat-on-flat system at rest does not increase for $t_w > \tau_\mathrm{MD}$ while the rough-on-flat system sees a significant increase in XSLs on time-scales longer than the relaxation time, mirroring the increase in $\Delta \mu_\mathrm{MD}$ shown in \cref{fig:observation}e. The difference between flat and rough systems can be explained with contact junctions that are free to evolve in the rough case, i.e.\ molecules at junction boundaries can accommodate contact stresses by moving their tail in/out of junctions, whereas this mechanism cannot occur if the entire interface participates in contact. The similarities between $\Delta N$ and $\Delta\mu_\mathrm{MD}$ prompts us to investigate the number of links in the second slide stage to understand how the system looses its memory of the contact interface and rejuvenates to a constant friction force.

\begin{figure}
\includegraphics{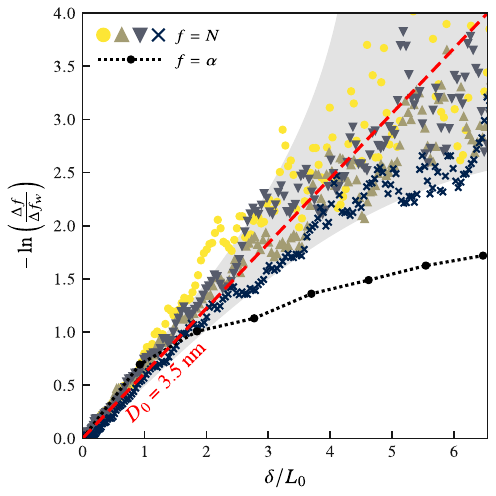}
\caption{\textbf{Evolution of the number of XSLs ($f = N$, symbols), and contact survival fraction ($f = \alpha$, dashed line with circles) in slide-after-hold as a function of the sliding distance $\delta$}. The dashed (red) line shows a decay of the form $f - f_{ss} = \Delta f \sim \exp(-\delta / D_0)$ with $D_0$ = \SI{3.5}{\nm}. This decay holds well for $N$ even for several times $D_0$, regardless of sliding velocity
(symbol shapes, all with $t_w/\tau_\mathrm{MD}$ = 300), suggesting that $N$ is a representative quantity of the interface state.
%($v \cdot \tau_\mathrm{MD} / L_0 =$ 0.7, 1.2, 1.4, 1.9 for the symbols \protect\tikz[scale=0.120pt]{\protect\fill[black] (0, 0) circle[radius=1];}, \protect\tikz[scale=0.25pt]{\protect\fill[black] (0, 0) -- ($(0.5, 0)+sqrt(3)/2*(0, 1)$) -- (1, 0);}, \protect\tikz[scale=0.25pt, rotate=180]{\protect\fill[black] (0, 0) -- ($(0.5, 0)+sqrt(3)/2*(0, 1)$) -- (1, 0);}, \protect\tikz[scale=0.20pt]{\protect\draw[thick] (0, 0) -- (1, 1) (1, 0) -- (0, 1);} respectively)
Dispersion of the data on long sliding distances is expected due to the natural noise of the systems, which introduces uncertainty in the steady-state estimate (gray area). The dry contact curve ($f = \alpha$) diverges from $N$, indicating that $D_0$ is not an exclusive property of the roughness.}\label{fig:bonds_rescaled}
\end{figure}

\section*{Transients governed by sub-junction mechanisms}
In \cref{fig:bonds_rescaled}, we note $\Delta f = f(\delta) - f_{ss}$ and $\Delta f_w = f_w - f_{ss}$ with $f$ being either the number $N$ of XSLs or the contact area survival fraction $\alpha$, $f_w$ and $f_{ss}$ denoting these quantities at the end of the holding stage and at steady-state, respectively. The contact area survival fraction $\alpha$ is measured from continuum simulations of frictionless elastic dry rough-on-rough contact where the top surface is shifted by $\delta$ and the true contact area $A(\delta)$ is compared to the initial ($\delta = 0$) contact area: $\alpha = |A(\delta) \cap A(0)| / |A(0)|$. The quantity $-\ln(\Delta f / \Delta f_w)$ gives an idea as to how $f$ returns to its steady state value, i.e. loses its memory, as the top surface slides a distance $\delta$: rate-and-state models that use the aging law $\dot{\phi} = 1 - v\phi/D_0$ predict that $\phi(\delta) - \phi_{ss} \propto \exp(-\delta / D_0)$. In \cref{fig:bonds_rescaled}, axes are chosen so that an exponential decay is a straight line with slope $1 / D_0$ in \cref{fig:bonds_rescaled}. Symbols show that the number of XSLs ($f = N$) for different sliding velocities ($v \cdot \tau_\mathrm{MD} / L_0 =$ 0.7, 1.2, 1.4, 1.9% in case we need symbols, below is the tikz code
%for the symbols \tikz[scale=0.120pt]{\fill[black] (0, 0) circle[radius=1];}, \tikz[scale=0.25pt]{\fill[black] (0, 0) -- ($(0.5, 0)+sqrt(3)/2*(0, 1)$) -- (1, 0);}, \tikz[scale=0.25pt, rotate=180]{\fill[black] (0, 0) -- ($(0.5, 0)+sqrt(3)/2*(0, 1)$) -- (1, 0);}, \tikz[scale=0.20pt]{\draw[thick] (0, 0) -- (1, 1) (1, 0) -- (0, 1);} respectively
) follows an exponential decay with $D_0$ = \SI{3.5}{\nm}. This is consistent, in magnitude, with the distance needed for the experimental friction force to return to steady-state, measured to be 4.8 $\pm$ \SI{1.4}{\nm}. Although the number of XSLs and the friction force should not be directly compared, the fact that the MD simulations reproduce a value of $D_0$ independent of sliding velocity and in the same order of magnitude as the experiment is a remarkable result, considering the 10 orders of magnitude difference in sliding velocity between simulations and experiments, and that no simulation parameter was adjusted to match the experimental data.

Uncertainty (due to noise) in the measurement of the steady-state value of an exponential decay causes deviations from the straight line. The gray area in \cref{fig:bonds_rescaled} shows the deviation extent based on the noise in $N$ measured in the simulations. Unlike $N$, the contact survival fraction, $\alpha$, for dry elastic contact (dotted line), decays to a steady-state at a slower rate than the other quantities, indicating that $D_0$ is not an intrinsic property of the junction sizes, as is commonly interpreted~~\citep{dieterichDirectObservationFrictional1994}, but rather a velocity-independent system property~\citep{molinariFundamentalAspectsNew2019,tianMemoryDistanceInterfacial2019} that combines the surface roughness and the molecular organization of the fatty acid molecules.
%To our knowledge, this is the first time that $D_0$ is tied to a physical mechanism that we show occurs \emph{within} contact junctions, and yet clearly influences the macroscopic friction response.

\begin{figure}[ht]
    \centering
    \includegraphics{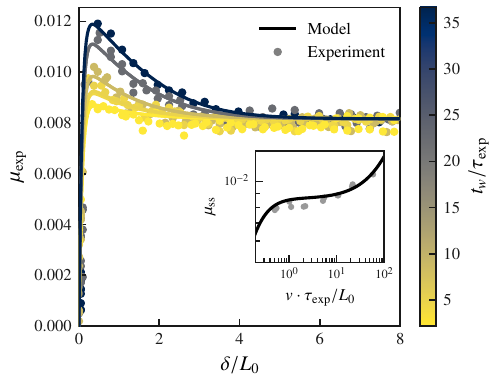}
    \caption{\textbf{Transient friction predicted by multi-scale friction model.} The theoretical prediction (black line) is compared to the experimental friction transient ($v$ = \SI{0.5}{\nm/\s}) and steady-state (inset) responses (circles). The characteristic times of the nano-scale contribution to the friction force are consistent with measured relaxation times (cf.~\cref{fig:observation}), and the model is capable of reproducing the transient friction behavior: the theoretical friction overshoots the steady-state value and decays over the same distance as the experimental results.}
    \label{fig:model}
\end{figure}

Our experiments and simulations show that taking the surface roughness into account, even at the nano scale, and by extent the formation of contact junctions, is a key component in the structural frictional aging, although junctions could arise from other sources of heterogeneity, like imperfect surface coverage of the adsorbed layer. We also demonstrate that the cross-surface link formation and destruction within contact junctions govern key aspects of the aging process and of the transient frictional response. We combine these two ideas into a model that links the nano and macroscopic scales---in both time and space---and analytically reproduces the steady-state friction response as well as the transient overshoot in the presence of roughness. This model unifies existing approaches at two length scales: the macro scale where the true contact area is made up of monolayer junctions due to the presence of surface roughness and the scale of molecular interactions within a junction. At this molecular scale, three characteristic times are defined: the time to break a molecular link (i.e.\ cross-surface link), the time to (re)activate a molecular link, and the delay time related to the withdrawal of a link from the interpenetration zone. These ingredients allow the modeling of friction both in stationary and transient regimes, while accounting for the time, sliding velocity, surface roughness, and elastic properties of the monolayers. The friction force $F_{t}(v,t)$ is decomposed as the product of an interfacial shear stress, $\sigma_s (v)$ (which depends on the sliding velocity), the real contact area, $A_r$, and a state term $f_\mathrm{a}(\phi)$~\citep{baumbergerPhysicalAnalysisState1999} (which depends on the age of the contacts $\phi(t)$ obeying the evolution equation $\dot{\phi} = 1 - v\phi/D_0$). The interfacial shear stress $\sigma_s$ is calculated according to the Chernyak--Leonov theory~\citep{chernyakTheoryAdhesiveFriction1986, leonovDependenceFrictionForce1990} using the characteristic times defined above. In the inset of \cref{fig:model}, we show our fit of $F_{t,ss}$ to the steady-state experimental values of the friction force at different sliding velocities (values of the model parameters are given in the methods section). For the characteristic detachment time, we find values in the same order of magnitude as the relaxation time measured in \cref{fig:observation}, and as values measured for different organic monolayers with the same thickness~\citep{corwinFrictionalAgingSliding2009}. To account for transient effects at the onset of sliding, we combine the described approach with an extension of Mindlin's model~\citep{mindlinComplianceElasticBodies1949, bureauElasticiteRheologieInterface2002} to multi-asperity contacts~\citep{greenwoodContactNominallyFlat1966}. The friction force evolution with the sliding distance is then expressed as $F_t(\delta,v,t) = F_{t}(v,t) \left[1-\exp(-\delta /\delta^*)\right]$ where $\delta^*$ is the ratio of the maximum friction force to the measured tangential stiffness of the interface (the hypothesis leading to the full derivation of this equation are given in the methods section) and illustrated in \cref{fig:model}. Without roughness, our model correctly predicts no overshoot of the stationary friction value, in agreement with our MD simulations. This demonstrates that the physics of friction is well captured by the complex coupling between the dynamics of link formation inside contact junctions, governed by the aforementioned characteristic times, and the sliding dynamics of the junctions themselves all over the contact area. Thus, the interface accommodates the shearing through a combined effect of roughness and molecular interdigitation. This approach can readily be generalized to other systems with adsorbed organic layers and rough surfaces, which are commonplace in microelectromechanical systems~\citep{corwinFrictionalAgingSliding2009} and biomechanics~\citep{jinBiofriction2013}, e.g.\ for natural or artificial joints where proteins can form a protective layer on a hard substrate~\citep{widmerInfluencePolymerSurface2001}. Finally, the methodology we described to combine nano and macro scales is generic and should be applicable to a wide variety of frictional systems, particularly in geophysics, e.g.\ for friction laws of glacier beds on rough topography~\citep{joughinRegularizedCoulombFriction2019,helanowSlipLawHardbedded2021}, which play a role in climate modeling~\citep{fox-kemperOceanCryosphereSea2021}.

\section*{Conclusions}
Combining experiments and simulations of friction between fatty acid monolayers deposited on rough surfaces, and complemented by a multi-scale theoretical approach, we were able to probe the molecular mechanisms underlying frictional aging and the transient friction response, and bridge the scale gap to the macroscopic friction behavior. We have shown, for monolayers adsorbed on rough surfaces, that the transient friction characteristic scale is controlled by molecular mechanisms within contact junctions, not the sizes of the junction themselves. This sheds new light on the memory length-scale $D_0$ with important implications on a broad range of frictional systems described by rate-and-state models. We have also demonstrated that in the absence of contact junctions (e.g. due to the absence of roughness), structural frictional aging disappears. Aging can then be explained by the capacity of molecules to come in and out of contact junctions, which cannot happen when surfaces are flat at the atomic scale. Our findings highlight the importance of both surface roughness at the molecular scale and molecular mechanisms at the macroscopic scale. We combined these aspects into a multi-scale theoretical model that correctly reproduces the transient friction overshoots observed in experiments, and whose principles are generalizable to a wide variety of dry and lubricated frictional systems with surface roughness, including geological faults, where the interplay between friction and junction size plays a role in the slip regime of the fault~\citep{lebihainEarthquakeNucleationFaults2021}, or biomechanical systems, e.g.\ cartilaginous or artificial joints, for which roughness can dramatically alter the proper function and lifetime~\citep{espinosaCartilageAssessmentRequires2021}.

% , with a unified predictive theoretical approach, we are able to finally bridge the gap and reconcile nano and macro scales to tackle rough friction. We show that the existence of the roughness, even nanometric, plays a major role in forming junctions whose dynamics govern the macroscopic friction. Space and time scales are coupled using two space scales, namely the accommodation distance, $D_0$, and the macroscopic contact size via $\delta^*$ and its dependence on the tangential stiffness, and using three times characteristic of molecular motions within the junction and one time corresponding to the age of the macroscopic contact. This allowed us to perfectly predict transient friction overshoots as well as the steady-state macroscopic response. Faut il ouvrir vers les perspectives de l'abstract?

%%%%%%%%%%%%%%%%%%%%%%%%%%%%%%%%%%%%%%%%%%%%%%%%%%%%%%%%%%%%%%%%%%%%%%%%%%%%%%%%

\bibliography{biblio}

\section*{Acknowledgements}
This article is dedicated to the memory of Mark O. Robbins.

\subsection*{Funding} LF acknowledges the financial support of the Swiss National Science Foundation (grant \#191720 ``Tribology of Polymers: from Atomistic to Continuum Scales'') and the Johns Hopkins University. LF and JAE acknowledge the computational support of the Advanced Research Computing at Hopkins (ARCH) core facility (rockfish.jhu.edu), which is supported by a U.S. National Science Foundation (NSF) grant number OAC-1920103. LF and JAE also acknowledge the computational support of the Extreme Science and Engineering Discovery Environment (XSEDE) Expanse supercomputer at the San Diego Supercomputer Center (SDSC) through allocation TG-MAT210003. XSEDE is supported by NSF grant number ACI-1548562. JAE acknowledges financial support from the NSF CAREER award CMMI-1454072. AC, DM and JCB acknowledge the financial support of the French National Research Agency (Confluence project ANR-13-JS09-0016-01), and of the Agency for the ecological transition (ADEME) through the IMOTEP project.

\subsection*{Author contributions} All authors participated in discussions. JCB and MOR initiated the collaboration. LF, JAE, DM and JCB wrote and edited the article. JCB and DM devised the experiments, AC produced the experimental data. LF, MOR and JAE devised the MD simulations, LF produced the simulation data. DM established the unified theoretical model.

\subsection*{Competing interests} The authors declare no competing interests.

\subsection*{List of supplementary materials}
\begin{itemize}
    \item Materials and Methods
    \item Supplementary figures S1-S6
\end{itemize}

\if\preprint1 % Submission setup
\pagebreak
\fi

\section*{Materials and Methods}
\subsection*{Experimental Friction Measurement}
Using stearic acid ($99.0~\%$ purity, from Sigma-Aldrich) with dehydrated and filtered dodecane, a dilute solution is prepared at a concentration of \SI{0.002}{\mol/\L}. The surfaces consist of a fused silicate glass sphere of radius 2.030 $\pm$ \SI{0.005}{\mm} and a $\langle 100 \rangle$ silicon wafer. The latter is cleaned with isopropanol and deionized water using a spin-coater at 8000 rpm and then dried under nitrogen flow. Both surfaces are then coated with a \SI{40}{\nm}-thin cobalt layer by means of a cathodic sputtering system under low argon pressure (\SI{e-6}{\milli\bar}). Experiments are conducted by sliding the sphere over the plane using the ATLAS molecular tribometer~\citep{crespoMethodologyCharacterizeRheology2017}. A typical friction experiment is performed by approaching the sphere towards the plane, confining the stearic acid monolayers on each surface until a constant normal force of 0.70 $\pm$ \SI{0.01}{\mN}, i.e.\ a corresponding maximum Hertzian contact pressure of \SI{27}{\MPa} (at a normal velocity of \SI{0.1}{\nm/\s}). Then without breaking contact, a slide-hold-slide procedure is used: the sphere slides over the plane over a few hundreds nanometers at a constant sliding velocity of \SI{0.5}{\nm/\s}, then is held stationary for a time, $t_w$, before resuming the lateral displacement. Hold times are varied between \SI{1}{\s} and \SI{120}{\s}. During the experiment, the response to a superimposed oscillating sphere displacement in both directions, normal and tangential, of amplitude \SI{0.1}{\nm} and \SI{38}{\Hz} (respectively \SI{0.03}{\nm} and \SI{70}{\Hz}) provides, without disturbing the friction process, the stiffness and the viscous damping of the confined interface in both directions~\citep{crespoMethodologyCharacterizeRheology2017}. All measurements are carried out in a sealed chamber of relative humidity lower than $1~\%$ and $T$ = $23.0^\circ$C $\pm$ $0.5^\circ$C in an argon atmosphere.

\subsection*{Surface Topography Characterization}
Multi-scale characterization of the surface topography is performed before and after the experiment to ensure no surface damage. AFM measurements of the surface topography over an area of \SI{1}{\um}$\times$\SI{1}{\um} provide an RMS of surface heights of \SI{0.6}{\nm} and a radially averaged power-spectrum density (PSD) shown in supplementary \cref{ex:psd_exp}. At a larger scale, Bruker interferometry profilometer provides an RMS value of 0.5 nm on both surfaces in Phase Shift Interferometry mode over an area of \SI{63}{\um}$\times$\SI{47}{\um}. 

To generate synthetic rough surfaces from the measured surface profile, we use the PSD computed from the AFM data. We cut off long wavelength modes as necessary to generate a smaller surface, i.e.\ for MD simulations that are \SI{200}{\nm}$\times$\SI{200}{\nm}, and use uniformly distributed phases~\citep{wuSimulationRoughSurfaces2000} to produce surfaces with the same (or reduced) spectral content as the surface used in experiments. Full size (\SI{1}{\um}$\times$\SI{1}{\um}) surfaces are used for continuum simulations of dry elastic contact, while reduced size (\SI{200}{\nm}$\times$\SI{200}{\nm}) are used for MD simulations.

\subsection*{Molecular Dynamics}
Molecular dynamics simulations are conducted using coarse-grained potentials~\citep{salernoResolvingDynamicProperties2016} adjusted for alkane chains with one bead corresponding to two CH\textsubscript{2} groups. Stearic acid chains consist of nine beads. Head groups are positioned on a hexagonal lattice with spacing \SI{5.5}{\angstrom}. The top lattice is rotated \SI{90}{\degree} to avoid commensurate effects in the flat/flat friction response. The applied normal pressure is $\bar{p}$ = \SI{27}{\MPa}. Roughness is applied to the head-group lattice by vertical displacement of the beads and their connected chain. The system is initially equilibrated at $T$ = \SI{300}{\kelvin}, with the surfaces separated, using a Langevin thermostat and a time-step of $\Delta t$ = \SI{1}{\fs}. Surfaces are brought together with the applied normal pressure and equilibrated again. Sliding of the top head-group lattice is done via a spring attached to its center of mass. The stiffness of the spring is such that the period of the mass-spring system is \SI{3.5}{\ps}. In the initial sliding phase, the free end of the spring slides at velocity $v$ for \SI{600}{\angstrom} and $\Delta t$ = \SI{1.25}{\fs}. The system is then allowed to rest by setting $v$ to zero for \SI{30}{\ns} with $\Delta t$ = \SI{3}{\fs}. Restart of the sliding is done by setting $v$ back to its original value with $\Delta t$ = \SI{1.25}{\fs}. The friction force is measured as the force in the spring, and the temperature is controlled at \SI{300}{\kelvin} with a Langevin thermostat acting on the degrees of freedom normal to the sliding direction. The number of cross-surface links is computed with a radius cutoff of \SI{10}{\angstrom} for attractive links and \SI{5}{\angstrom} for compressive links, which correspond to the potential cutoff and equilibrium distance respectively. All simulations are conducted with the open-source software \href{https://www.lammps.org}{LAMMPS}~\citep{plimptonFastParallelAlgorithms1995,thompsonLAMMPSFlexibleSimulation2022}.

\subsection*{Continuum Elastic Rough Contact}
A Fourier-based boundary integral approach~\citep{stanleyFFTBasedMethodRough1997,frerotFourieracceleratedVolumeIntegral2019} is used with a projected conjugate gradient algorithm~\citep{polonskyNumericalMethodSolving1999} to solve the elastic rough contact problem. The linear elastic material properties used are determined from the experiments~\citep{crespoEffectUnsaturationAdsorption2018}: the contact Young's modulus $E^* = E/(1-\nu^2)$ is set to \SI{48}{\GPa} and the average pressure is set to $\bar{p}$ = \SI{27}{\MPa}. The contact problem is solved with a compound roughness~\citep{johnsonContactMechanics1985} $h = h_2 - h_1$ from two generated surfaces $h_{1}$ and $h_2$, the latter of which is shifted by $\delta$, the sliding distance. The true contact area is the area where contact pressure is strictly positive. The survival fraction at $\delta$ is the normalized magnitude of the area in common with the initial contact area. All simulations are conducted with the open-source library \href{https://gitlab.com/tamaas/tamaas}{\textsc{Tamaas}}~\citep{frerotTamaasLibraryElasticplastic2020,frerotTamaasHighperformanceLibrary2022}.

\subsection*{Junction-based Friction Model}
The coupling between the roughness scale of the multi-asperity contact and the molecular scale at which the two stearic acid monolayers interact is considered (see \cref{fig:observation}). This interaction only occurs in contact junctions between surface asperities. The evolution of friction is then modeled at these two scales~\citep{crespoComprehensionTribologieFilms2017}.

\paragraph{In the steady-state regime---} The friction force $F_{t}$ is assumed to admit the decomposition $F_t(v,t) = \sigma_s(v)\cdot A_r f_a\left[\phi(t)\right]$, where the interfacial shear strength $\sigma_s(v)$ is given by the Chernyak--Leonov's model~\citep{chernyakTheoryAdhesiveFriction1986}, $A_r / A_\mathrm{Hertz} = 0.04$ determined from Ref.~\citep{pastewkaContactAreaRough2016} for a soft wall repulsion, and the age factor $f_\mathrm{a} = 1+\omega\ln\left(1+\phi/\tau_1\right)${~\citep{bureauElasticiteRheologieInterface2002}} depends on the age variable $\phi$ following $\dot{\phi} = 1 - v\phi/D_0$. $D_0$ in the non-stationary evolution equation acts as a characteristic memory length-scale: at the molecular scale, it is the sliding distance needed to renew the cross-surface links, as shown in the simulations. The Chernyak--Leonov theory is used to describe the dynamics of links formation between the stearic acid molecules, through the interpenetration zone according to three elementary times ~\citep{leonovDependenceFrictionForce1990}: $\tau_\mathrm{0}$, the time necessary to break a link, $\tau$, the time necessary to form a link and $\hat{\tau}$ the time for a molecule to withdraw from the interpenetration zone. According to this model, the interfacial shear strength $\sigma_s$ can be written as:
$$\sigma_s(v) = \sigma_0 u\frac{1-(1+m+1/u)\exp\left(-m-1/u\right)}{1+m\gamma-\exp\left(-m-1/u\right)}$$
with $u= \tan\chi \cdot v\tau_0 / (2L_0)$, $m=\tau/\tau_\mathrm{0}$, $\gamma=\tau/\hat{\tau}$, and $\sigma_\mathrm{0}=\left(2G/\tan\chi\right)\cdot\left(L_\mathrm{0}/L_\mathrm{H}\right)$ deduced from  Ref.~\citenum{chernyakTheoryAdhesiveFriction1986} where $\chi$
is the angle made by the stretched molecule in sliding. In the expression of $\sigma_s$, only $\tau_0$, $\gamma$ and $\chi$ are free parameters: $m$ can be computed using \cref{ex:msqd}, while $G$, $L_H$ are measured and $L_0$ can be found in the literature~\citep{askwithChainLengthAdditives1966}. The coefficients $\omega$ and $\tau_1$ in $f_\mathrm{a}$ can be obtained from~\cref{fig:observation}. We find that both $\tau_1$ and $\tau_0$ have values in the same order of magnitude as $\tau_\mathrm{exp}$ = \SI{2.2}{\s} and values reported in the literature for different organic compounds but similar monolayer thickness~\citep{corwinFrictionalAgingSliding2009}, confirming that they relate to a chain relaxation mechanism as postulated by \citet{chernyakTheoryAdhesiveFriction1986}.

\paragraph{Onset of sliding---} The macroscopic transient friction force $F_t$ for an interface transitioning from rest to a sliding velocity $v$ is calculated from the extension of Mindlin's theory to rough surfaces~\citep{bureauElasticiteRheologieInterface2002}. This is possible due to the sliding distance $\delta$ being much smaller than the characteristic diameter of the contact junctions (see supplementary \cref{ex:contact_map_reslide}). In a multi-asperities interface, this means some contact spots remains in partial sliding while others are moved in total sliding with a friction coefficient $\mu(v,t)=F_\mathrm{t}(v,t)/F_\mathrm{n}$ where $F_\mathrm{t}(v,t)$ is given in the previous paragraph. The critical sliding distance, $\delta^* = F_{t,ss} / K_x$ experimentally measured as \SI{0.16}{\nm} for $v$ = \SI{0.5}{\nm/\s} is introduced. It is interpreted as the distance required to switch from partial to total sliding~\citep{mazuyerFrictionDynamicsConfined2008}. Thus, according to the Mindlin theory, the elementary tangential force $f$ required to move a micro-contact in partial sliding is simply $f=\mu(v,t) f_\mathrm{n}\cdot\left[1-\left(1-\delta/\delta^*\right)^{3/2}\right]$, where $f_\mathrm{n}$ is the load applied on one micro-contact, with the assumption that all contacts have the same age $\phi$. The Greenwood-Williamson model applied to such a multi-contact interface ~\citep{bureauElasticiteRheologieInterface2002} gives $F_t(\delta, v, t)=F_\mathrm{t}(v,t)\left[1-\exp\left(-\delta/\delta^*\right)\right]$.

The model uses values of $G$, $\delta^{*}$, and $L_\mathrm{H}$ measured during the experiments, $D_0$ evaluated from $D_\mathrm{ss}$, the distance necessary for the friction force to decay to steady state and $m$ evaluated in MD relaxation simulations. The nanoscale parameters governing the steady-state response are $\tau_0$, $\gamma$ and $\chi$, which are adjusted independently of $\omega$ and $\tau_1$. The values used in \cref{fig:model} are listed in \cref{tab:data}. We find that $\gamma$ is effectively zero, indicating that the retraction time $\hat{\tau}$ is very large compared to the attachment time $\tau$.
 
\sisetup{separate-uncertainty}
\begin{table}[htbp]
\centering 
\caption{Numerical values used for the junction-based friction model.} \label{tab:data}
\begin{tabular}{l@{\hspace{0.5cm}}ll}
\toprule
Measured values & Steady-state param.  & Aging param.\\
\midrule
$L_\mathrm{0}$ = \SI{2.14}{\nm} & $\tau_0$ = \SI{5.37+-0.89}{\s} & $\tau_1$ = \SI{5}{\s}\\
$L_\mathrm{H}$ = \SI{1.55+-0.2}{\nm} & $\gamma$ = \num{0.0} & $\omega$ = \num{0.278}\\
$G$ = \SI{6.3+-1.2}{\MPa} & $\chi$ = \ang{55.3+-1.6}\\
$\delta^{*}$ = \SI{0.16}{\nm} \\
$D_\mathrm{ss}$ = \SI{4.8+-1.4}{\nm} \\
$m$ = \num{2.5e-3} \\
$K_x$ = \SI{43}{\kN\m^{-1}}\\
\bottomrule
\end{tabular}
\end{table}

To summarize, this theory combines the main friction models developed in tribology in the past eight decades, each model representing one of the contributions to friction physics:
\begin{itemize}
\item Greenwood--Williamson model that describes the effect of elastically deformed roughness;
\item Mindlin theory that proposes a physical process of static/kinetic friction based upon the evolution of the adhesive part of a macro or micro-contact;
\item Chernyak--Leonov approach, derived from Schallamach's model~\citep{schallamachVelocityTemperatureDependence1953}, that predicts the effect of the dynamics of molecular links breaking/formation during sliding;
\item Adhesive Bowden--Tabor law that highlights the role of interfacial shear strength of junctions and the real contact area;
\item State and rate phenomenology revisited according to \citet{baumbergerPhysicalAnalysisState1999} to take into account the contact aging and rejuvenation of a multi contact interface.
\end{itemize}

% \par\noindent\rule{\columnwidth}{0.4pt}

\pagebreak

\makeatletter
\setcounter{figure}{0}
\renewcommand{\thefigure}{S\@arabic\c@figure}
\newcounter{SIfig}
\renewcommand{\theSIfig}{S\arabic{SIfig}}
\crefalias{SIfig}{figure}
\makeatletter

\begin{figure*}
    \centering
    \includegraphics{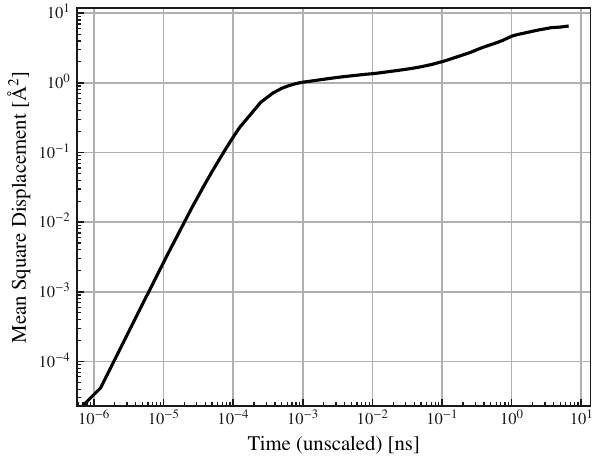}
    \refstepcounter{SIfig}\label{ex:msqd}
    \caption{\textbf{Mean square displacement of beads (MD), with respect to unscaled time.} The time scaling factor, $\alpha$ for the coarse-grained model used~\citep{salernoResolvingDynamicProperties2016} is 7.63. The first characteristic timescale corresponds to the end of the ballistic regime ($t/\alpha\approx$  \SI{e-3}{\ns}) and gives the monomer relaxation time. The second characteristic timescale corresponds to the chain relaxation timescale, at $t/\alpha\approx$ \SI{e-1}{\ns} $\Rightarrow \tau_\mathrm{MD}$ = \SI{0.8}{\ns}.}
\end{figure*}

\begin{figure*}
    \centering
    \includegraphics{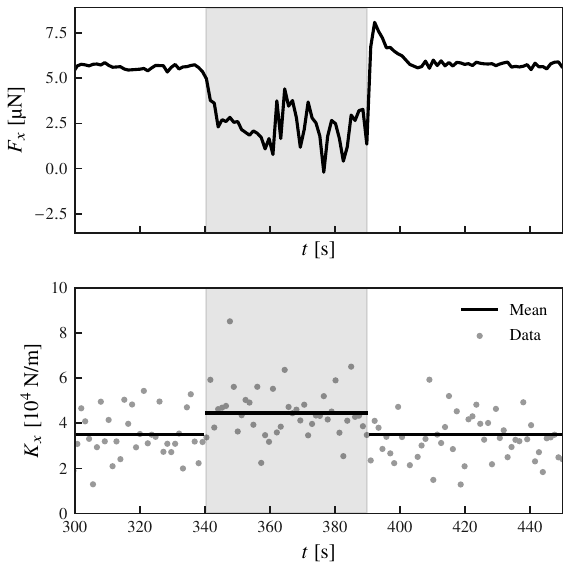}
    \refstepcounter{SIfig}\label{ex:tangential_stiffness}
    \caption{\textbf{Structural aging of the interface (experimental).} Simultaneous evolution of the friction force (top) and the tangential stiffness (bottom) during an SHS process. A slight increase in the tangential stiffness, $K_x$, is measured during the holding stage, which spans \SI{50}{\s}. No variation of the film thickness is detected. ~\citep{crespoComprehensionTribologieFilms2017}}
\end{figure*}

\begin{figure*}
    \centering
    \includegraphics{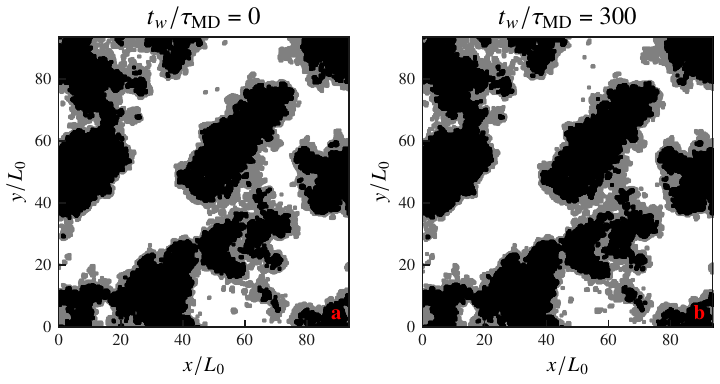}
    \refstepcounter{SIfig}\label{ex:contact_map_wait}
    \caption{\textbf{Contact area evolution during the holding stage (MD).} Black zones indicate repulsive interaction, while gray zones indicate attractive interaction. \textbf{a} shows the state of the contact area before the holding stage (at the end of the sliding stage); while figure \textbf{b} shows the end of the holding stage for $t_w/\tau_\mathrm{MD}$ = 300. No significant difference can be observed.}
\end{figure*}

\begin{figure*}
    \centering
    \includegraphics{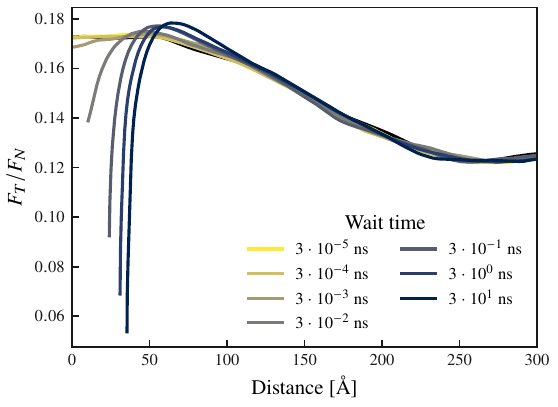}
    \refstepcounter{SIfig}\label{ex:rough_static_peaks}
    \caption{\textbf{Friction overshoot peaks in a rough-on-rough system (MD).} Due to the roughness present on both surfaces, the steady-state friction in not constant, but varies as the contact interface changes in the sliding process. Wait time shown are in unscaled units.}
\end{figure*}

\begin{figure*}
    \centering
    \includegraphics{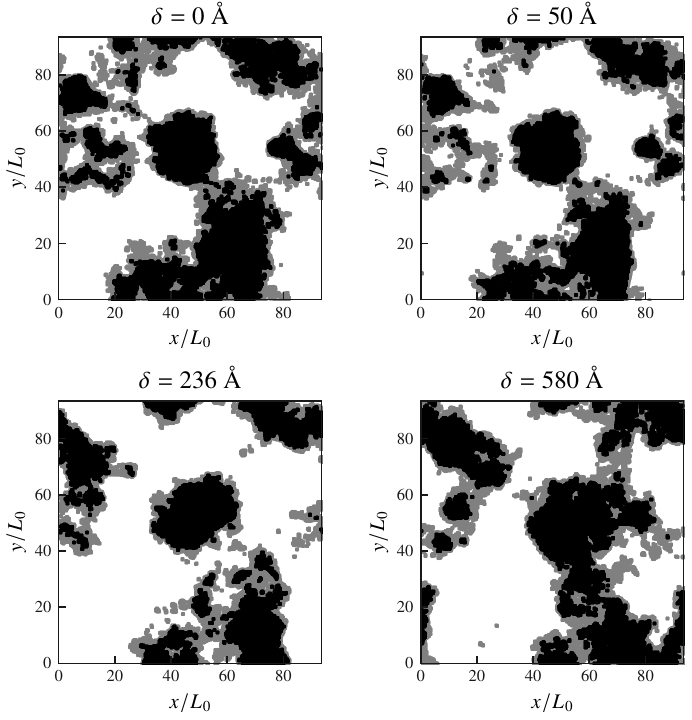}
    \refstepcounter{SIfig}\label{ex:contact_map_reslide}
    \caption{\textbf{Contact area evolution in the second sliding stage of the rough-on-rough system (MD).} Each image corresponds to a different value of the sliding distance, $\delta$. On distances of the order of the characteristic scale $D_0$ = \SI{3.5}{\nm}, the true contact area is not significantly modified by the shifting of the two rough surfaces.}
\end{figure*}

\begin{figure*}
\centering
\includegraphics{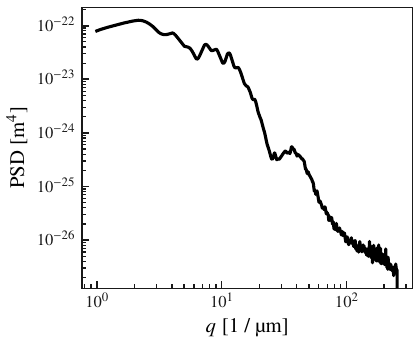}
\refstepcounter{SIfig}\label{ex:psd_exp}
\caption{\textbf{Power-spectrum density of tribometer surface topography.} Topography was measured with AFM over an area of \SI{1}{\um}$\times$\SI{1}{\um}.}
\end{figure*}

\end{document}